\documentclass[conference]{IEEEtran}
\ifCLASSINFOpdf
   \usepackage[pdftex]{graphicx}
\else
\fi

\usepackage{flushend}
\usepackage{multirow}

\begin{document}
%
\title{Multi-Parameter Decision Support with Data
Transmission over GSM/GPRS Network: a Case
Study of Landslide Monitoring}

\author{\IEEEauthorblockN{Satyajit Rath, B. P. S. Sahoo, S. K. Pandey and D. P. Sandha}
\IEEEauthorblockA{Computer Networking \& e-Management Division\\
CSIR-Institute of Minerals and Materials Technology\\
Bhubaneswar, Odisha, India - 751 013\\
E-mail: satya@immt.res.in}
}


%


\maketitle

\begin{abstract}
The planet Earth has hundreds of impact events,
with some occurrences causing both in terms of human casualty
as well as economic losses. Such attitudes of earth pushed the
frontiers to develop innovative monitoring strategies for the earth
system. To make that real, although, will require coherent and
real-time data by observing the earth behaviour contiguously.
Wireless Sensor Network (WSN) appears to be the best suitable
infrastructure to sense environmental parameters of our interests.
In this event of earth observation, another important issue is the
monitoring system with high level of precision. There are different
types of sensors to measure the behavioural aspects of earth. The
sensors integrated with WSN, provide an accurate and contiguous
data for analysis and interpretation. This paper briefly addresses
earth observation and areas of critical importance to people
and society. A case study has also been carried out for disaster
like Landslide in the North Eastern region of India. Application
software has been developed for the said study for online data
acquisition and analysis with pre-disaster early warning system.
The system monitors the changing geo-technical condition of this
region using various geo-technical sensors like Rain gauge, In-
place Inclinometer, Tilt-meter, Piezo-meter and Crack meter.
This paper also touches upon the aspects of data transmission
over Global System for Mobile Communication (GSM) / General
Packet Radio Service (GPRS) to a remote data center.
\end{abstract}


%
\IEEEpeerreviewmaketitle

\section{Introduction}
Disaster prevention and mitigation are integral to development activities. Major natural disasters can and do have
sever negative short-run economic impacts. Natural disasters
cause significant loss both in terms of human casualty as
well as budgetary development. In the recent years, the death
and damage in natural disasters have increased. Hence, in
this context there is good reason to study and analyse the
sensational aspects of disasters. There may have some general
principles which creates this irregularity which results in
disasters. In this irregularity, whether some general principles
exist or not, the answer may be ’yes’. Therefore it is necessary
to gather information about earth’s physical, chemical and
biological changes in a global scale.
\par
India is being affected by hydro-geological hazards like
landslide since years ago. History depicts that, it has happened
mostly in Himalayan regions, Hill ranges of Northeast, Western \& Eastern Ghats of India. The North East (NE) region of
India is badly affected by bewildering varieties of landslides;
specifically the state of Arunachal Pradesh. The entire hilly
region of the state is geologically very young and unstable.
This make the landslide monitoring as an important aspects
and challenges for the geo-scientists in India. Most of the
landslides in India are primarily caused due to heavy rainfall.
Depending on meteorological and physiographical conditions,
individual rainfall events can cause slope failures in areas of
limited extent or in large regions. Developing an early warning
system for the monitoring of landslides requires its domain
expertise, not just to build the instruments but to use them
properly and interpret their output for rational purpose.
\par
Knowing the occurrences of this event is a difficult task
to tackle and requires a thorough study of past activities of
the geo-technical condition as well as the local calamities.
However, this event can possibly be predicted if proper geo-
technical investigations of the triggering parameters are per-
formed. Currently, no system exists for NE region of India
to identify the triggering of landslides, largely due to lack
of observing the changing conditions of the geology at this
region. Apart from the landslides, the state is also prone to
various other hazards such as earthquake, floods, forest fires,
cyclones, etc. Due to high seismicity, high annual rainfall and
geological fragility of the region, almost all parts of the state
experience one or more of these disasters every year \cite{arun}.
\par
In this paper, we proposed a framework to develop a real-
time monitoring and prediction system looking at the causing
parameters around NE region of India with global approach.
This paper first articulates the real-time rainfall observations to
assess landslide hazards. The goal of this system is to acquire
a global view rather than the site specific view to design
a fully functional application software module for landslide
monitoring. A pilot study has been made at the deployment
site to identify the causing parameters and subsequently their
corresponding thresholds from the past event that has resulted
in landslides.

\section{Site Investigation}
In this section we will discuss the location of site where the proposed monitoring base station is to be place and the
geo-technical condition of this area. The loss and damage due to this life-threatening event for a period of three years i.e. from 2008 to 2011 is summarized also.

\subsection{Site Location and Geo-technical Conditions}
The site selected for monitoring purpose is located in Karshingsa, on the National Highway (NH) 52(A) to Itanagar, capital of Arunachal Pradesh. A long distance of road from Banderdewa to Naharlagun, the worst affected road section of
this NH, runs on the hill with river Dikrong flowing in one side. This NH52(A) is only the suitable communication way
to communicate with the neighbouring state Assam. The traffic movement to Itanagar gets badly affected by any disruption to the this highway.
\par
The NE region of India falls under the category of very high and high hazardous area in Landslide Hazard Zones in India \cite{lodhi}. The India Govt. has found five landslides sites viz. Powari, Nathpa Landslide, Himachal Pradesh; Sherka Danda Slide, Nainital, Uttar Pradesh; Chanmari, Sikkim; Karsingsa, Arunachal Pradesh, for detailed geological and geo-technical investigations. Karsingsa hill region comprises of some Hillocks, which are composed of some unique form of soil successions with the river Dikrong flowing just adjacent to the NH. Most of the landslides in Karsingsa occur every year during the month of June-July and normally triggered by the toe erosion at river bank due to action of rain and river water of Dikrong \cite{jotisankasa}. An in-depth geo-technical study has been performed to cognize the soil structure and gradient of the deployment site. As per the study, the grade of the soil structure is very unstable and weak; hence prone to slack.

\subsection{Damage and Loss Perspective}
A large scale and long duration landslide of continuous nature was occurred during 1998 which created havoc and loss
of life and property and interrupted the surface communication for a period of three months. Many places within the capital complex near Karsingsa sinking zone, Lekhi village, A-sector of Naharlagun, Papunallah area has also been severely affected by the landslide. These prototype events demand the need for methods to have an alarming system for rainfall-induced landslides at Karsingsa. Table \ref{table:loss} lists a few incidences of this life-threatening event in this area.

\begin{table}
\begin{center}
\caption{Landslide Incidences in Arunachal Pradesh, India}
    \begin{tabular}{|c||c||c|}
    \hline
    \bf{TIME}       & \bf{LOSS}     & \bf{CAUSE}          \\ \hline
    14/06/2008 & 15 death & Heavy rainfall \\ \hline
    25/07/2008 & 20 death & Heavy rainfall \\ \hline
    30/07/2009 & Unknown & 3 days continuous rain            \\ \hline
    02/04/2010 & 7 death  & 7 days continuous rain             \\ \hline
    21/04/2010 & 12 death & Heavy rainfall \\ \hline
    04/10/2010 & 10 death & Heavy rainfall \\ \hline
    22/11/2010 & Unknown  & Heavy rainfall \\ \hline
    16/07/2011 & Unknown  & Heavy rainfall \\ \hline
    16/08/2011 & Unknown  & Heavy rainfall \\ \hline
    \multicolumn{3}{|c| }{Source: Survey of India}\\ \cline{1-3}
    \end{tabular}
\end{center}
\end{table}

\subsection{Sensor Identified for Monitoring}
The detailed geo-technical field investigation reveals, the rainfall, as the most influencing factor for the occurrence of
the Landslide at this area. The study also enjoins that the pore-water pressure as an essential ingredient for accurate
prediction of onset landslides. It has been shown from sixteen years of direct observation in lower Himalayan region pertaining to Arunachal Pradesh; in general, most of the landslide occurred due to heavy rainfall \cite{jotisankasa}. The geo-technical survey and further detailed investigations shows the importance of soil displacement measure and the inclination of the steep along with the rainfall and pore-water pressure records, for more accuracy of detecting the event. To end with, the geo-technical investigation and the geo-physically condition of the deployment site suggests four parameters for the purpose of monitoring and an instrumentation setup with wireless network to gather the required parameter values through sensor devices. Data collected from the instrumentation setup are transmitted to the base station for further analysis. The parameters to be monitored and corresponding sensors are the following:
\begin{itemize}
  \item Rainfall intensity (Rain Gauge)
  \item Pore water pressure (Piezometer)
  \item Displacement of rock masses (Extensometer)
  \item Inclination of natural slopes (Inclinometer / Tilt meter)
\end{itemize}
To monitor the behaviour of the soil deformation round the clock, above five numbers of sensors are installed at the site.
In-place inclinometer, tiltmeter and piezometer are grouted in the boreholes, where as other two are placed on its appropriate position above the soil to measure the changing effects. Rain-gauge measures the rainfall intensity in millimetres that occurs over a unit area. In-place inclinometer and tiltmeter determines the inclination of the steep with the movements and deformation of the soil mass. The piezometer observes the pore water pressure using vibrating wires and the soil displacement is measured with Crack meter.
\begin{figure}[h!]
    \centering
    \includegraphics[width=0.48\textwidth]{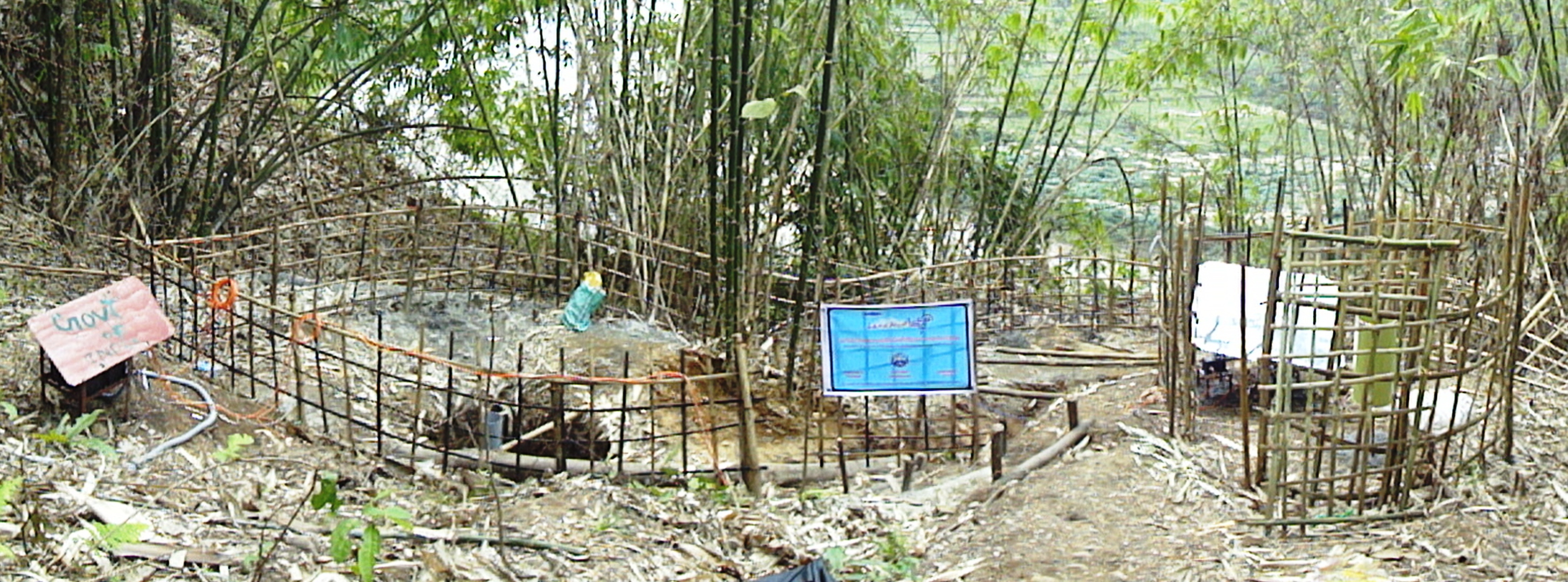}
\caption{Sensor node deployed at site.}
  \label{fig:sensite}
\end{figure}
\par
The figure \ref{fig:sensite} shows deployment site where the above sensor
nodes are grouted or deployed.

\section{Related Works}
The researchers around the globe are working to find out a potential solution to address this high economic and natural
loss due to landslides. So far wireless sensor network is the best suited technology to deal with these sensitive and risk areas \cite{akyildiz} \cite{sahoo2012}. Paper \cite{yang} reports a real implementation and deployment system for monitoring in the Northern Italy Apennines; and analyses the problems and inherent limitations they had faced in developing and deploying such a system. In the similar line, \cite{zan} deployed a wireless sensor network based landslide detection system at Idukki, the Southern state of Kerala, India. An effective method for data collection and aggregation by implementing threshold alert levels has been developed in \cite{maneesha}. A monitoring system based on National
Instrument LabView software and an A/D (analogue/digital) converter with internal processor, geophones, pressure transducer and a rain gauge has been prepared by the authors in \cite{fausto}. Jotisankasa and Hunsachainan in \cite{jotisankasa} reports on the development of an early warning system for landslide monitoring for the site at Thadan dam, Nakhonnayok province, Thailand based on the values of rainfall, pore water pressure and shear strains. A framework for developing an experimental real-time prediction system to identify the rainfall-triggered landslides has been proposed in \cite{lodhi} by combining the surface landslide susceptibility and a real-time space-based rainfall analysis system. Ken Tsutsui et al has presented a new technique to detect large-scale landslides as well as to estimate their volume based on the elevation changes of digital elevation models extracted from high resolution satellite stereo imagery in \cite{richard}.
The authors in \cite{xianguo} discussed the problem of landslide in mining area and propose a landslide prediction model based on support vector machine. A spatial data mining has been introduced to make landslide spatial prediction; and a suitable spatial data mining algorithm for landslide prediction has been proposed in \cite{gaelle} based on cloud model and data field.
\par
The study does not illustrate a monitoring system which addresses every aspects of the monitoring system. Most of the
work is targeting to the initial data accumulation and store. In this contribution, an approach of data communication over
General Packet Radio Systems (GPRS) network, data analysis and event prediction is clearly presented for a complete system prototype.

\section{Sensor Data Transmission over GPRS Network}
This section describes the detail architecture of the data transmission from deployment site to local monitoring station.
The collected sensor data is sent to the local station over wireless medium. The data repository server at the local station
receives the transmitted sensor data and stored them in a database server for further analysis.
\subsection{Technology Used}
As mentioned the data is being transfer over a wireless medium, we found the GPRS technology can be the one, among other available technologies for mobile data transfer. GPRS is a non-voice value added service that allows voice and data to be sent and received across a mobile telephone network. Though it is already a proved technology for data transfer for which we do not need new infrastructures. We can build the application on the existing technology and resources available to us.
\par
The data rates of GPRS is up to 115 kbps \cite{sahooiitb}, which would be well supported for our amount of data to be transmitted. The objective of the adoption of this technology is to reduce the effort to build new wireless infrastructure and investment. The most important advantages of this technology used here is, we can place the monitoring station at any place in any corner as per the availability of GPRS network. Hence this is most suited wireless data transmission technology which significantly meet our purpose.
\par
To enable this above said technology we have used GSM/GPRS modem at each end and developed a algorithmic module for encoding and decoding before transmitting and receiving respectively.
\subsection{Deployment Site GPRS Unit}
The system is designed by ensuring self configuration and communication over the GPRS network. At the deployment
site, the sensors units are connected to a micro controller board (MCB) with proper power connection from solar panel. An
GPRS modem is also placed on the the MCB. When the this MCB unit boots, it initiates a signal to read/write to the GPRS
modem. Immediately then, MCB requests the Internet Service Provider (ISP) for an Internet Protocol (IP) address by sending a ’REQIP’ packet. The micro controller waits until the modem is not assigned with an IP. Soon after an IP is being assigned to the modem, micro controller sends this IP to the monitoring station via a text message and requests for server IP by sending ’SENDIP’ packet. After receiving the IP of data repository server it requests for a connection by sending ’REQCONN’ packet. After the connection establishment, the MCB sends the sensor data by heading ’SENDDATA’ packet. During the transmission of sensor data, the MCB also checks the connection status. If it founds the connection has been broke then it re-establish the connection and repeat the sequence again.
\subsection{Monitoring Station GPRS Unit}
At the base station, an GPRS modem (here after called server-modem) is connected to RS232 port of the data-server
system. An application module always reads the RS232 port for a new connection to establish. When it receives the text
message from client-modem it extract the IP address from the text message add into the client list. Then on receiving
the ’SENDIP’ packet from the client-modem, it sends his own IP address to the client-modem. Then it receives a
connection request from the client-modem, and then accept the connection request for data transmission purpose. Finally,
after the connection establishment it receives sensor data by recognizing ’SENDDATA’ packet. The technology used in this work is influenced from the work presented in \cite{sahooiitb}, where author used GPS, GSM technology for location tracking and monitoring.
\par
This connection initiation, establishment, re-establishment and sending data are executed with the help of pre-specified
AT commands for GPRS modem.

\section{Decision Support System based Data Analysis and Monitoring}
Data collected from the sensor nodes are transmitted over GPRS network to the base station and being stored in a
relational database for further analysis. The network infrastructure for landslide monitoring is generally consists of a data
acquisition and a data distribution network, which is monitored and controlled by a base station \cite{richard}.
\par
As the objective of the system is to alert the local people and authorities, an early warning system is developed. To predict
the sensor data, we have used a multi-parameter based model as shown in the Fig. \ref{fig:model}.
\subsection{Sensor Data Analysis and Prediction}
\begin{figure}[h!]
    \centering
    \includegraphics[width=0.45\textwidth]{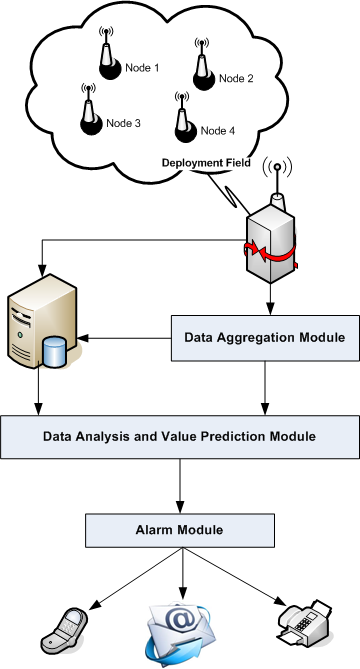}
\caption{Decision Support Model for Event Prediction and Early Warning.}
  \label{fig:model}
\end{figure}

The mechanism for landslide monitoring, prediction and information dissemination about the occurrence of the event has
been studied in-depth. As understood from the literature survey, no multi-dimensional and integrated mechanism has been found to fulfil all the said aspects of landslides monitoring. In this pilot scale of study, a novel methodology for the detection, prediction and information dissemination for landslides has been developed using machine learning technique Artificial Neural Network (ANN) and Multi Criteria Decision Analysis (MCDA) model. Fig. \ref{fig:model}, shows the designed working model for our extensive study for data analysis, prediction and early warning.
\par
The data aggregation module calibrates the raw data with the given error constants of each sensor before storing the data to the database. Based on the hardware specification of the sensors and calibration, the error constants for each sensor were identified. Finally, the data aggregation module stores the formatted data in the data repository for further analysis.
Here onwards the model plays a significant role to raise the alarm. As it can be seen from the Fig \ref{fig:model}, the data from the aggregation module is being sent to the data repository for maintaining the historical data as well as to the application system for analysis. The designed system generates alerts based on the current sensor values in uni-parameter based and multi-parameter based analysis. However this mechanism does not provide much time to the executants to go for the remedial measures. Looking at the need, a prediction mechanism was envisaged and implemented. This prediction
mechanism uses the historical data from the data repository and provides the input for both uni-parameter based \& multi-parameter based analysis. It is prudent to mention here that, the system is capable of generating alarms in four ways i.e. with the current sensor values and also with predicted sensor values in combination with uni-parameter based and multi-parameter based. The Value Prediction Model uses machine learning technique such as ANN and/or Auto Regression (AR) to predict each sensor value from its past occurrences in the landslides.

\par
It has been experienced, that the mechanism of slope failure at our pilot site is basically a static liquefaction and rainfall is
recognized as one of the most highly valued triggering factor. The geo-technical investigation suggests that the groundwater conditions is also responsible for slope failures and are related to rainfall through infiltration. Occurrence of Landslide in response to rainfall involves physical processes that operate on disparate timescales. For rainfall-induced landslides, a threshold may define the rainfall, soil moisture, or hydrological conditions that when reached or exceeded, are likely to trigger landslides \cite{richard}. In 1980, Nel Caine was first to propose a global rainfall intensity duration threshold for the occurrence of shallow landslides and debris flows \cite{yang}. Using local precipitation records, he defined an upper threshold for landslide initiation as presented in Equation \ref{eq:one}.
\begin{equation}
I = 14.82~*~D^{-0.39}(0.167 < D < 500)
\label{eq:one}
\end{equation}
where, I is rainfall intensity in millimetre per hour and D is rainfall duration in hours.
\par
A statistical threshold for the occurrence of landslide has been measured based on the analysis of historical inventory of
landslides events. It has been defined by investigating various parameters like: total rainfall; antecedent rainfall; rainfall
intensity, and rainfall duration.

\subsection{Multi-Parameter Level Early Warning}
In our study we have adopted a multi level warning system to give alert for the onset landslides, which has not been found
in the literature so far. There are two aspects to generate the alarm such as with current sensor value and predicted sensor value with uni-parameter based and multi-parameter based. In uni-parameter based, the alert gets triggered when any one of the influencing factor reaches or exceeds its threshold value where as in muti-level based it considers multiple factors for the threshold. These alerts also can be triggered on the predicted sensor value and current sensor value. However, the system follows the same mechanism to generate the alarm. The system compares the sensor values (current/predicted) with the monitoring threshold (MT) and works in the following manner:

\begin{itemize}
\item Green Signal Warning or No Warning: When all the captured sensor value are less than MT, during this period the system only shows a green button representing no current risk or everything is in normal situation. This level is only for the monitoring purpose at the base station. In this situation no warning will be given to the local authority or local people.
\item Yellow Signal or First Level Warning: When the most valued triggering factor i.e. rainfall intensity reaches or exceeds the MT(Rain Gauge) value, then the system shows a orange button in the developed monitoring software interface representing that there is a chance of occurrence of landslide in the recent future. During this period a warning is given through beeping sound at the monitoring station and text message / fax to the local authority. 
\item Orange Signal or Second Level Warning: The rainfall induces increase in water table / the pore water pressure. Subsequently this triggers the next level of warning. When pore pressure reaches or exceeds the MT (Piezometer) along with MT (Rain Gauge), the system shows an ORANGE button representing further chance of occurrence of landslide in recent future. This also sends alert SMS to concerned authorities. This level of alert suggests avoiding going to that region.
\item Red Signal or Third Level Warning: Due to increase in pore water pressure, the landmass starts displacing from its original location and in turn triggers the landslide. When other triggering factor i.e. displacement of masses / inclination of natural slopes reaches or exceeds the MT(Extensometer) / MT(Inclinometer)  respectively along with MT (Rain Gauge) and MT (Piezometer), the system shows a RED button representing a greater chance of occurrence of landslide in recent future. During this level local people to be alerted about the occurrence of landslide by louder speaker or sirens. This also sends alert SMS to concerned authorities. This level of alert suggests the local people to vacate the place or avoid going to that region.
\end{itemize}

\section{Conclusion}
This paper presents a novel methodology to monitor the occurrence of landslides and predict the same before it happens.
Based on the geo-technical circumstances of the deployment site, the possible required parameters were duly selected.
Beyond the analysis at local base station, this paper en-lights the possibility of data transmission over a GSM/GPRS network to a remote site. This in turn will help to monitor multiple base stations placed at far distances. The adopted methodology gives the provision to disseminate the captured data for further analysis and reporting purpose.
\par
This shows the possible ways and means of detecting the any kind of natural disasters by observing the responsible parameters. The software has been developed with its dynamism and can be configured to address any type of sensor based monitoring needs.

\section*{Acknowledgment}
This research is supported and funded by the Ministry of Communication \& Information Technology (MCIT), Department of Electronics and Information Technology (DeitY), Govt. of India. The authors are thankful to Shri B. M. Baveja, Group Coordinator, CC\&BT, DeitY, MCIT, New Delhi. Authors are also thankful to the Directors and Scientists of respective CSIRs institutes for their kind guidance and support.

\end{document}